\documentclass[aps,pra,twocolumn, groupedaddress,tightenlines]{revtex4}

\usepackage{amsmath}
\usepackage[english]{babel}

\begin{document}

\title{Robustness of operator quantum error correction with respect to initialization errors}

\author{Ognyan Oreshkov}

\affiliation{Department of Physics, Center for Quantum Information
Science \& Technology, University of Southern California, Los
Angeles, California 90089, USA}

\date{\today}

\begin{abstract}

In the theory of operator quantum error correction (OQEC), the
notion of correctability is defined under the assumption that
states are perfectly initialized inside a particular subspace, a
factor of which (a subsystem) contains the protected information.
If the initial state of the system does not belong entirely to the
subspace in question, the restriction of the state to the
otherwise correctable subsystem may not remain invariant after the
application of noise and error correction. It is known that in the
case of decoherence-free subspaces and subsystems (DFSs) the
condition for perfect unitary evolution inside the code imposes
more restrictive conditions on the noise process if one allows
imperfect initialization. It was believed that these conditions
are necessary if DFSs are to be able to protect imperfectly
encoded states from subsequent errors. By a similar argument,
general OQEC codes would also require more restrictive
error-correction conditions for the case of imperfect
initialization. In this study, we examine this requirement by
looking at the errors on the encoded state. In order to
quantitatively analyze the errors in an OQEC code, we introduce a
measure of the fidelity between the encoded information in two
states for the case of subsystem encoding. A major part of the
paper concerns the definition of the measure and the derivation of
its properties. In contrast to what was previously believed, we
obtain that more restrictive conditions are not necessary neither
for DFSs nor for general OQEC codes. This is because the effective
noise that can arise inside the code as a result of imperfect
initialization is such that it can only increase the fidelity of
an imperfectly encoded state with a perfectly encoded one.

\end{abstract}

\maketitle


\section{Introduction}

Operator quantum error correction (OQEC) \cite{KLP05,KLPL06,BKK07}
is a generalized approach to protecting quantum information from
noise, which unifies in a common framework previously proposed
error correction schemes, including the standard method of active
error correction \cite{Sho95,Ste96,BDSW96,KL96} as well as the
passive method of decoherence-free subspaces
\cite{DG98,ZR97,LCW98,LBKW01} and subsystems
\cite{KLV00,DeF00,KBLW01,YGB01} (for a recent generalization
including entanglement-assisted error correction, see
\cite{HDB07,GHWAC07}). This approach employs the most general
encoding for the protection of information---encoding in
subsystems of the Hilbert space of a system \cite{Knill06} (see
also Ref.~\cite{KNPV07}). The concept of noiseless subsystem is a
cornerstone in this theory, as it serves as a basis for the
definition of correctable subsystem and error correction in
general. This concept is defined through the assumption of perfect
initialization of the state of the system inside a particular
subspace. In practice, however, perfect initialization of the
state may not be easy to achieve. Hence, it is important to
understand to what extent the preparation requirement can be
relaxed.

As shown in Ref.~\cite{ShaLid05}, in order to ensure perfect
noiselessness of a subsystem in the case of imperfect
initialization, the noise process has to satisfy more restrictive
conditions than those required in the case of perfect
initialization. It was believed that these conditions are
necessary if a noiseless (or more generally decoherence-free)
subsystem is to be robust against arbitrarily large initialization
errors. The fundamental relation between a noiseless subsystem and
a correctable subsystem implies that in the case of imperfect
initialization, more restrictive conditions would be needed for
OQEC codes as well.

In this paper we show that with respect to the ability of a code
to protect from errors, more restrictive conditions are not
necessary. For this purpose, we define a measure of the fidelity
between the encoded information in two states for the case of
subsystem encoding. We first give an intuitive motivation for the
definition, and then study the properties of the measure. We then
show that the effective noise that can arise inside the code due
to imperfect initialization under the standard conditions, is such
that it can only increase the fidelity of the encoded information
with the information encoded in a perfectly prepared state. This
robustness against initialization errors is shown to hold also
when the state is subject to encoded operations.

\section{Review of conditions for noiseless subsystems and OQEC
codes}

For simplicity, we consider the case where information is stored
in only one subsystem, i.e., we consider a decomposition of the
system's Hilbert space of the form
\begin{equation}
\mathcal{H^S}=\mathcal{H^A}\otimes\mathcal{H}^B\oplus
\mathcal{K},\label{decomposition}
\end{equation}
where the sector $\mathcal{H}^A$ (also called a subsystem) is used
for encoding of the protected information. Let
$\mathcal{B}(\mathcal{H})$ denote the set of linear operators on a
finite-dimensional Hilbert space $\mathcal{H}$. In the OQEC
formalism, noise is represented by a completely positive
trace-preserving (CPTP) linear map or a noise channel
$\mathcal{E}:\mathcal{B}(\mathcal{H}^S)\rightarrow
\mathcal{B}(\mathcal{H}^S)$. Every such map can be written in the
Kraus form \cite{Kraus83}
\begin{equation}
\mathcal{E}(\sigma)=\underset{i}{\sum}E_i\sigma E_i^{\dagger},
\hspace{0.2cm} \textrm{for all } \sigma\in
\mathcal{B}(\mathcal{H}^S),
\end{equation}
 where the Kraus operators
$\{E_i\}\subseteq\mathcal{B}(\mathcal{H}^S)$ satisfy
\begin{equation}
\underset{i}{\sum}E_i^{\dagger}E_i=I^S.\label{completeness}
\end{equation}

The subsystem $\mathcal{H}^A$ in the decomposition
\eqref{decomposition} is called noiseless with respect to the
channel $\mathcal{E}$, if
\begin{gather}
\textrm{Tr}_B\{(\mathcal{P}^{AB}\circ\mathcal{E})(\sigma)\}=\textrm{Tr}_B\{\sigma\},\label{noiselesssystem}\\
\hspace{0.1cm} \textrm{for all }\sigma\in
\mathcal{B}(\mathcal{H}^S) \textrm{ such that }
\sigma=\mathcal{P}^{AB}(\sigma)\hspace{0.1cm} ,\notag
\end{gather}
where $\mathcal{P}^{AB}(\cdot)=P^{AB}(\cdot)P^{AB}$ and $P^{AB}$
is the projector of $\mathcal{H}^S$ onto $\mathcal{H}^A\otimes
\mathcal{H}^B$ ($P^{AB}\mathcal{H}^S=\mathcal{H}^A\otimes
\mathcal{H}^B$). Similarly, a correctable subsystem is one for
which there exists a correcting CPTP map
$\mathcal{R}:\mathcal{B}(\mathcal{H}^S)\rightarrow
\mathcal{B}(\mathcal{H}^S)$, such that the subsystem is noiseless
with respect to the map $\mathcal{R}\circ \mathcal{E}$:
\begin{gather}
\textrm{Tr}_B\{(\mathcal{P}^{AB}\circ\mathcal{R}\circ\mathcal{E})(\sigma)\}=\textrm{Tr}_B\{\sigma\},\label{correctablesystem}\\
\hspace{0.1cm} \textrm{for all }\sigma\in
\mathcal{B}(\mathcal{H}^S) \textrm{ such that }
\sigma=\mathcal{P}^{AB}(\sigma)\hspace{0.1cm} .\notag
\end{gather}

The definition of noiseless subsystem \eqref{noiselesssystem}
implies that the information encoded in
$\mathcal{B}(\mathcal{H}^A)$ remains invariant after the process
$\mathcal{E}$, if the initial density operator of the system
$\rho(0)$ belongs to
$\mathcal{B}(\mathcal{H}^A\otimes\mathcal{H}^B)$. If, however, one
allows imperfect initialization,
$\rho(0)\neq\mathcal{P}^{AB}(\rho(0))$, this need not be the case.
Consider the ``initialization-free" analogue of the definition
\eqref{noiselesssystem}:
\begin{gather}
\textrm{Tr}_B\{(\mathcal{P}^{AB}\circ\mathcal{E})(\sigma)\}=\textrm{Tr}_B\{\mathcal{P}^{AB}(\sigma)\},\label{IFnoiselesssystem}\\
\hspace{0.1cm} \textrm{for all } \sigma\in
\mathcal{B}(\mathcal{H}^S).\notag
\end{gather}
Obviously Eq.~\eqref{IFnoiselesssystem} implies
Eq.~\eqref{noiselesssystem}, but the reverse is not true. As shown
in \cite{ShaLid05}, the definition \eqref{IFnoiselesssystem}
imposes more restrictive conditions on the channel $\mathcal{E}$
than those imposed by \eqref{noiselesssystem}. To see this,
consider the form of the Kraus operators $E_i$ in the block basis
corresponding to the decomposition \eqref{decomposition}. From a
result derived in \cite{ShaLid05} it follows that the subsystem
$\mathcal{H}^A$ is noiseless in the sense of
Eq.~\eqref{noiselesssystem}, if and only if the Kraus operators
have the form
\begin{equation}
E_i=\begin{bmatrix} I^A\otimes C_i^B&D_i\\
0&G_i
\end{bmatrix},\label{Krausoperatorsblock}
\end{equation}
where the upper left block corresponds to the subspace
$\mathcal{H}^A\otimes\mathcal{H}^B$, and the lower right block
corresponds to $\mathcal{K}$. The completeness relation
\eqref{completeness} implies the following conditions on the
operators $C^B_i$, $D_i$, and $G_i$:
\begin{gather}
\underset{i}{\sum}C_i^{\dagger B}C_i^B=I^B,\label{one}\\
\underset{i}{\sum}I^A\otimes C_i^{\dagger B}D_i=0,\label{two}\\
\underset{i}{\sum}(D_i^{\dagger}D_i+G_i^{\dagger}G_i)=I_{\mathcal{K}}\label{three}.
\end{gather}

In the same block basis, a perfectly initialized state $\rho$ and
its image under the map \eqref{Krausoperatorsblock} have the form
\begin{gather}
{\rho}=\begin{bmatrix} {\rho}_1&0\\
0&0
\end{bmatrix}, \hspace{0.2cm} \mathcal{E}(\rho)=\begin{bmatrix}
\rho_1'&0\\
0&0
\end{bmatrix},\label{perfectini}
\end{gather}
where $\rho_1'=\underset{i}{\sum}I^A\otimes C_i^B\rho_1 I^A\otimes
{C_i^{\dagger B}}$. Using the linearity and cyclic invariance of
the trace together with Eq.~\eqref{one}, we obtain
\begin{gather}
\textrm{Tr}_B\{(\mathcal{P}^{AB}\circ\mathcal{E})(\rho)
\}=\textrm{Tr}_B\{\underset{i}{\sum}I^A\otimes C_i^B\rho_1
I^A\otimes {C_i^{\dagger B}}\}\notag\\=\textrm{Tr}_B\{\rho_1
\underbrace{\underset{i}{\sum}I^A\otimes C_i^{\dagger
B}C_i^B}_{I^A\otimes
I^B}\}=\textrm{Tr}_B\{\mathcal{P}^{AB}(\rho)\},\label{reducedrho}
\end{gather}
i.e., the reduced operator on $\mathcal{H}^A$ remains invariant.

On the other hand, an imperfectly initialized state $\tilde{\rho}$
and its image have the form
\begin{equation}
\tilde{\rho}=\begin{bmatrix} \tilde{\rho}_1&\tilde{\rho}_2\\
\tilde{\rho}_2^{\dagger}&\tilde{\rho}_3
\end{bmatrix}, \hspace{0.2cm} \mathcal{E}(\tilde{\rho})=\begin{bmatrix}
\tilde{\rho}_1'&\tilde{\rho}_2'\\
\tilde{\rho}_2'^{\dagger}&\tilde{\rho}_3'
\end{bmatrix}.\label{imperfect}
\end{equation}
Here $\tilde{\rho}_2$ and/or $\tilde{\rho}_3$ are non-vanishing,
and
\begin{gather}
\tilde{\rho}_1'=\underset{i}{\sum}(I^A\otimes C_i^B\tilde{\rho}_1
I^A\otimes {C_i^{\dagger B}}+D_i\tilde{\rho}_2^{\dagger}I^A\otimes
{C^{\dagger B}_i}\\+I^A\otimes C^B_i\tilde{\rho}_2
D_i^{\dagger}+D_i\tilde{\rho}_3D_i^{\dagger}),\notag\\
\tilde{\rho}_2'=\underset{i}{\sum}(I^A\otimes C_i^B\tilde{\rho}_2
G_i^{\dagger}+D_i\tilde{\rho}_3G_i^{\dagger}),\\
\tilde{\rho}_3'=\underset{i}{\sum}G_i\tilde{\rho}_3G_i^{\dagger}.
\end{gather}
In this case, using the linearity and cyclic invariance of the
trace together with Eq.~\eqref{one} and Eq.~\eqref{two}, we obtain
\begin{widetext}
\begin{gather}
\textrm{Tr}_B\{(\mathcal{P}^{AB}\circ\mathcal{E})(\tilde{\rho})
\}=\textrm{Tr}_B\{\underset{i}{\sum}(I^A\otimes
C_i^B\tilde{\rho}_1 I^A\otimes {C_i^{\dagger B}}
+D_i\tilde{\rho}_2^{\dagger}I^A\otimes {C^{\dagger
B}_i}\notag+I^A\otimes C^B_i\tilde{\rho}_2
D_i^{\dagger}+D_i\tilde{\rho}_3D_i^{\dagger})\}
\notag\\=\textrm{Tr}_B\{\tilde{\rho}_1
\underbrace{\underset{i}{\sum}I^A\otimes C_i^{\dagger
B}C_i^B}_{I^A\otimes
I^B}\}+\textrm{Tr}_B\{(\underbrace{\underset{i}{\sum}{I^A\otimes
C^{\dagger B}_i}D_i}_{0})\tilde{\rho}_2^{\dagger}
\}+\textrm{Tr}_B\{\tilde{\rho}_2(\underbrace{\underset{i}{\sum}{I^A\otimes
C^{\dagger
B}_i}D_i}_{0})^{\dagger}\}+\textrm{Tr}_B\{\underset{i}{\sum}D_i\tilde{\rho}_3D_i^{\dagger}
\}\notag\\
=\textrm{Tr}_B\tilde{\rho}_1+\textrm{Tr}_B\{\underset{i}{\sum}D_i\tilde{\rho}_3D_i^{\dagger}
\}\neq
\textrm{Tr}_B\tilde{\rho}_1\equiv\textrm{Tr}_B\{\mathcal{P}^{AB}(\tilde{\rho})\},\label{reducedrhotilde}
\end{gather}
\end{widetext}
i.e., the reduced operator on $\mathcal{H}^A$ is not preserved. It
is easy to see that the reduced operator would be preserved for
every imperfectly initialized state if and only if we impose the
additional condition
\begin{equation}
D_i=0, \hspace{0.2cm} \textrm{for all } i.\label{extraconstraint}
\end{equation}
This further restriction to the form of the Kraus operators is
equivalent to the requirement that there are no transitions from
the subspace $\mathcal{K}$ to the subspace
$\mathcal{H}^A\otimes\mathcal{H}^B$ under the process
$\mathcal{E}$. This is in addition to the requirement that no
states leave $\mathcal{H}^A\otimes\mathcal{H}^B$, which is ensured
by the vanishing lower left blocks of the Kraus operators
\eqref{Krausoperatorsblock}. Condition \eqref{extraconstraint}
automatically imposes an additional restriction on the
error-correction conditions, since if $\mathcal{R}$ is an
error-correcting map in this ``initialization-free" sense, the map
$\mathcal{R}\circ \mathcal{E}$ would have to satisfy
Eq.~\eqref{extraconstraint}. But is this constraint necessary from
the point of view of the ability of the code to correct further
errors?

Notice that since $\tilde{\rho}$ is a positive operator,
$\tilde{\rho_3}$ is positive, and hence
$\textrm{Tr}_B\{\underset{i}{\sum}D_i\tilde{\rho}_3D_i^{\dagger}
\}$ is positive. The reduced operator on subsystem
$\mathcal{H}^A$, although unnormalized, can be regarded as a
(partial) probability mixture of states on $\mathcal{H}^A$. The
noise process modifies the original mixture
($\textrm{Tr}_B\tilde{\rho}_1$) by \textit{adding} to it another
partial mixture (the positive operator
$\textrm{Tr}_B\{\underset{i}{\sum}D_i\tilde{\rho}_3D_i^{\dagger}
\}$). Since the weight of any state already present in the mixture
can only increase by this process, this should not worsen the
faithfulness with which information is encoded in $\tilde{\rho}$.
In order to make this argument rigorous, however, we need a
measure that quantifies the faithfulness of the encoding.

\section{Fidelity between the encoded information
in two states}

\subsection{Motivating the definition}

If we consider two states with density operators $\tau$ and
$\upsilon$, a good measure of the faithfulness with which one
state represents the other is given by the fidelity between the
states:
\begin{equation}
F(\tau,
\upsilon)=\textrm{Tr}\sqrt{\sqrt{\tau}\upsilon\sqrt{\tau}}.\label{fidelity}
\end{equation}
This quantity can be thought of as a square root of a generalized
``transition probability" between the two states $\tau$ and
$\upsilon$ as defined by Uhlmann \cite{Uhl76}. Another
interpretation due to Fuchs \cite{Fuchs96} gives an operational
meaning of the fidelity as the minimal overlap between the
probability distributions generated by all possible generalized
measurements on the states:
\begin{equation}
F(\tau,
\upsilon)=\underset{\{M_i\}}{\textrm{min}}\underset{i}{\sum}\sqrt{\textrm{Tr}\{
M_i\tau\}}\sqrt{\textrm{Tr}\{ M_i\upsilon\}}.\label{fidelityoper}
\end{equation}
Here, minimum is taken over all positive operators $\{M_i\}$ that
form a positive operator-valued measure (POVM) \cite{Kraus83},
$\underset{i}{\sum}M_i=I^S$.

In our case, we need a quantity that compares the \textit{encoded}
information in two states. Clearly, the standard fidelity between
the states will not do since it measures the similarity between
the states on the entire Hilbert space. The encoded information,
however, concerns only the reduced operators on subsystem
$\mathcal{H}^A$. In view of this, we propose the following

\textit{Definition 1.} Let $\tau$ and $\upsilon$ be two density
operators on a Hilbert space $\mathcal{H}^S$ with decomposition
\eqref{decomposition}. The fidelity between the information
encoded in subsystem $\mathcal{H}^A$ in the two states is given
by:
\begin{equation}
F^A(\tau, \upsilon)=\underset{\tau',
\upsilon'}{\textrm{max}}F(\tau',\upsilon'),\label{measure}
\end{equation}
where maximum is taken over all density operators $\tau'$ and
$\upsilon'$ that have the same reduced operators on
$\mathcal{H}^A$ as $\tau$ and $\upsilon$: $\textrm{Tr}_B\{
\mathcal{P}^{AB}(\tau') \} = \textrm{Tr}_B\{
\mathcal{P}^{AB}(\tau) \}$, $\textrm{Tr}_B\{
\mathcal{P}^{AB}(\upsilon') \} = \textrm{Tr}_B\{
\mathcal{P}^{AB}(\upsilon) \}$.

The intuition behind this definition is that by maximizing over
all states that have the same reduced operators on $\mathcal{H}^A$
as the states being compared, we ensure that the measure does not
penalize for differences between the states that are not due
specifically to differences between the reduced operators.

\subsection{Properties of the measure}

\textit{Property 1 (Symmetry).} Since the fidelity is symmetric
with respect to its inputs, it is obvious from Eq.~\eqref{measure}
that $F^A$ is also symmetric:
\begin{equation}
F^A(\tau, \upsilon)=F^A(\upsilon, \tau).
\end{equation}

Although intuitive, the definition \eqref{measure} does not allow
for a simple calculation of $F^A$. We now derive an equivalent
form for $F^A$, which is simple and easy to compute. Let
$\mathcal{P}_{\mathcal{K}}(\cdot)=P_\mathcal{K}(\cdot)P_\mathcal{K}$
denote the superoperator projector on $\mathcal{B}(\mathcal{K})$,
and let
\begin{gather}
\rho^A\equiv\textrm{Tr}_B\{\mathcal{P}^{AB}(\rho)\}/\textrm{Tr}\{
\mathcal{P}^{AB}(\rho)\}
\end{gather}
denote the \textit{normalized} reduced operator of $\rho$ on
$\mathcal{H}^A$.

\textit{Theorem 1.} The definition \eqref{measure} is equivalent
to
\begin{eqnarray}
F^A(\tau, \upsilon)=f^A(\tau,\upsilon)\label{easyform}
+\sqrt{\textrm{Tr}\{
\mathcal{P}_{\mathcal{K}}(\tau)\}\textrm{Tr}\{
\mathcal{P}_{\mathcal{K}}(\upsilon)\}},
\end{eqnarray}
where
\begin{gather}
f^A(\tau,\upsilon)=\sqrt{\textrm{Tr}\{
\mathcal{P}^{AB}(\tau)\}\textrm{Tr}\{
\mathcal{P}^{AB}(\upsilon)\}}F(\tau^A,
\upsilon^A).\label{intermsofF}
\end{gather}

\textit{Proof.} Let $\tau^*$ and $\upsilon^*$ be two states for
which the maximum on the right-hand side of Eq.~\eqref{measure} is
attained. From the monotonicity of the standard fidelity under
CPTP maps \cite{BCF96} it follows that
\begin{gather}
F^A(\tau, \upsilon)=F(\tau^*, \upsilon^*)\leq F(\Pi(\tau^*),
\Pi(\upsilon^*)),
\end{gather}
where
$\Pi(\cdot)=\mathcal{P}^{AB}(\cdot)+\mathcal{P}_{\mathcal{K}}(\cdot)$.
But the states $\Pi(\tau^*)$ and $\Pi(\upsilon^*)$ satisfy
\begin{gather}
\textrm{Tr}_B\{ \mathcal{P}^{AB}(\Pi(\tau^*)) \} = \textrm{Tr}_B\{
\mathcal{P}^{AB}(\tau) \},\label{edno}\\
 \textrm{Tr}_B\{
\mathcal{P}^{AB}(\Pi(\tau^*))\} = \textrm{Tr}_B\{
\mathcal{P}^{AB}(\upsilon) \}\label{dve},
\end{gather}
i.e., they are among those states over which the maximum in
Eq.~\eqref{measure} is taken. Therefore,
\begin{gather}
F^A(\tau, \upsilon)= F(\Pi(\tau^*),
\Pi(\upsilon^*)).\label{pistar}
\end{gather}
Using Eq.~\eqref{fidelity} and the fact that in the block basis
corresponding to the decomposition \eqref{decomposition} the
states $\Pi(\tau^*)$ and $\Pi(\upsilon^*)$ have block-diagonal
forms, it is easy to see that
\begin{gather}
F(\Pi(\tau^*), \Pi(\upsilon^*))=
\check{F}(\mathcal{P}^{AB}(\tau^*), \mathcal{P}^{AB}(\upsilon^*))
\notag\\+ \check{F}(\mathcal{P}_{\mathcal{K}}(\tau^*),
\mathcal{P}_{\mathcal{K}}(\upsilon^*)),\label{intermediate}
\end{gather}
where $\check{F}$ is a function that has the same expression as
the fidelity \eqref{fidelity}, but is defined over all positive
operators. From Eq.~\eqref{edno} and Eq.~\eqref{dve} it can be
seen that $\textrm{Tr}\{ \mathcal{P}^{AB}(\tau^*)\}=\textrm{Tr}\{
\mathcal{P}^{AB}(\tau)\}$, $\textrm{Tr}\{
\mathcal{P}^{AB}(\upsilon^*)\}=\textrm{Tr}\{
\mathcal{P}^{AB}(\upsilon)\}$, which also implies that
$\textrm{Tr}\{ \mathcal{P}_{\mathcal{K}}(\tau^*)\}=\textrm{Tr}\{
\mathcal{P}_{\mathcal{K}}(\tau)\}=1-\textrm{Tr}\{
\mathcal{P}^{AB}(\tau)\}$, $\textrm{Tr}\{
\mathcal{P}_{\mathcal{K}}(\upsilon^*)\}=\textrm{Tr}\{
\mathcal{P}_{\mathcal{K}}(\upsilon)\}=1-\textrm{Tr}\{
\mathcal{P}^{AB}(\upsilon)\}$. The two terms on the right-hand
side of Eq.~\eqref{intermediate} can therefore be written as
\begin{gather}
\check{F}(\mathcal{P}^{AB}(\tau^*), \mathcal{P}^{AB}(\upsilon^*))=
\sqrt{\textrm{Tr}\{ \mathcal{P}^{AB}(\tau)\}\textrm{Tr}\{
\mathcal{P}^{AB}(\upsilon)\}}\notag\\\times
F\left(\frac{\mathcal{P}^{AB}(\tau^*)}{\textrm{Tr}\{
\mathcal{P}^{AB}(\tau)\}},
\frac{\mathcal{P}^{AB}(\upsilon^*)}{\textrm{Tr}\{
\mathcal{P}^{AB}(\upsilon)\}}\right),\label{firstterm}
\end{gather}
\begin{gather}
\check{F}(\mathcal{P}_{\mathcal{K}}(\tau^*),
\mathcal{P}_{\mathcal{K}}(\upsilon^*))= \sqrt{\textrm{Tr}\{
\mathcal{P}_{\mathcal{K}}(\tau)\}\textrm{Tr}\{
\mathcal{P}_{\mathcal{K}}(\upsilon)\}}\notag\\\times
F\left(\frac{\mathcal{P}_{\mathcal{K}}(\tau^*)}{\textrm{Tr}\{
\mathcal{P}_{\mathcal{K}}(\tau)\}},
\frac{\mathcal{P}_{\mathcal{K}}(\upsilon^*)}{\textrm{Tr}\{
\mathcal{P}_{\mathcal{K}}(\upsilon)\}}\right).\label{secondterm}
\end{gather}
Since $\tau^*$ and $\sigma^*$ should maximize the right-hand side
of Eq.~\eqref{intermediate}, and the only restriction on
$\mathcal{P}_{\mathcal{K}}(\tau^*)$ and
$\mathcal{P}_{\mathcal{K}}(\upsilon^*)$ is $\textrm{Tr}\{
\mathcal{P}_{\mathcal{K}}(\tau^*)\}=\textrm{Tr}\{
\mathcal{P}_{\mathcal{K}}(\tau)\}$, $\textrm{Tr}\{
\mathcal{P}_{\mathcal{K}}(\upsilon^*)\}=\textrm{Tr}\{
\mathcal{P}_{\mathcal{K}}(\upsilon)\}$, we must have
\begin{gather}
F\left(\frac{\mathcal{P}_{\mathcal{K}}(\tau^*)}{\textrm{Tr}\{
\mathcal{P}_{\mathcal{K}}(\tau)\}},
\frac{\mathcal{P}_{\mathcal{K}}(\upsilon^*)}{\textrm{Tr}\{
\mathcal{P}_{\mathcal{K}}(\upsilon)\}}\right)=1,
\end{gather}
i.e.,
\begin{gather}
\frac{\mathcal{P}_{\mathcal{K}}(\tau^*)}{\textrm{Tr}\{
\mathcal{P}_{\mathcal{K}}(\tau)\}} =
\frac{\mathcal{P}_{\mathcal{K}}(\upsilon^*)}{\textrm{Tr}\{
\mathcal{P}_{\mathcal{K}}(\upsilon)\}}.\label{tauupsstar}
\end{gather}
Thus we obtain
\begin{eqnarray}
\check{F}(\mathcal{P}_{\mathcal{K}}(\tau^*),
\mathcal{P}_{\mathcal{K}}(\upsilon^*))= \sqrt{\textrm{Tr}\{
\mathcal{P}_{\mathcal{K}}(\tau)\}\textrm{Tr}\{
\mathcal{P}_{\mathcal{K}}(\upsilon)\}}.
\end{eqnarray}
The term \eqref{firstterm} also must be maximized. Applying again
the monotonicity of the fidelity under CPTP maps for the map
$\Gamma(\rho^{AB})=\textrm{Tr}_B\{\rho^{AB}\}\otimes
|0^B\rangle\langle 0^B|$ defined on operators over
$\mathcal{H}^A\otimes\mathcal{H}^B$, where $|0^B\rangle$ is some
state in $\mathcal{H}^B$, we see that the term \eqref{firstterm}
must be equal to
\begin{gather}
\check{F}(\mathcal{P}^{AB}(\tau^*),
\mathcal{P}^{AB}(\upsilon^*))=\sqrt{\textrm{Tr}\{
\mathcal{P}^{AB}(\tau)\}\textrm{Tr}\{
\mathcal{P}^{AB}(\upsilon)\}}\notag\\ \times
F(\tau^A,\upsilon^A)\equiv f^A(\tau, \upsilon).\label{tauupsstar2}
\end{gather}
This completes the proof.

We next provide an operational interpretation of the measure
$F^A$. For this we need the following

\textit{Lemma.} The function $f^A(\tau,\upsilon)$ defined in
Eq.~\eqref{intermsofF} equals the minimum overlap between the
statistical distributions generated by all local measurements on
subsystem $\mathcal{H}^A$:
\begin{equation}
f^A(\tau,
\upsilon)=\underset{\{M_i\}}{\textrm{min}}\underset{i}{\sum}\sqrt{\textrm{Tr}\{
M_i\tau\}}\sqrt{\textrm{Tr}\{ M_i\upsilon\}},\label{fA}
\end{equation}
where $M_i=M_i^A\otimes I^B$, $\underset{i}{\sum}M_i=I^A\otimes
I^B$, $M^A_i>0$, for all $i$.

Note that since the operators $M_i$ do not form a complete POVM on
the entire Hilbert space, the probability distributions
$p_{\tau}(i)=\textrm{Tr}\{ M_i \tau\}$ and
$p_{\upsilon}(i)=\textrm{Tr}\{ M_i \upsilon\}$ generated by such
measurements generally do not sum up to 1. This reflects the fact
that a measurement on subsystem $\mathcal{H}^A$ requires a
projection onto the subspace $\mathcal{H}^A\otimes \mathcal{H}^B$,
i.e., it is realized through post-selection.

\textit{Proof.} Using that
\begin{gather}
\textrm{Tr}\{ M_i \tau\} =\textrm{Tr}\{M^A_i\otimes I^B
\mathcal{P}^{AB}(\tau)\}\notag\\=\textrm{Tr}\{
\mathcal{P}^{AB}(\tau)\}\textrm{Tr}\{M^A_i\otimes I^B
\frac{\mathcal{P}^{AB}(\tau)}{\textrm{Tr}\{
\mathcal{P}^{AB}(\tau)\}}\}\notag\\
=\textrm{Tr}\{ \mathcal{P}^{AB}(\tau)\}\textrm{Tr}\{M^A_i\tau^A\},
\end{gather}
we can write Eq.~\eqref{fA} in the form
\begin{gather}
f^A(\tau,\upsilon)=\sqrt{\textrm{Tr}\{
\mathcal{P}^{AB}(\tau)\}\textrm{Tr}\{
\mathcal{P}^{AB}(\upsilon)\}}\times\notag\\
\underset{\{M^A_i\}}{\textrm{min}}\underset{i}{\sum}\sqrt{\textrm{Tr}\{
M^A_i\tau^A\}}\sqrt{\textrm{Tr}\{
M^A_i\upsilon^A\}}.\label{interm}
\end{gather}
From Eq.~\eqref{fidelityoper}, we see that \eqref{interm} is
equivalent to \eqref{intermsofF}.

\textit{Theorem 2.} $F^A(\tau,\upsilon)$ equals the minimum
overlap
\begin{equation}
F^A(\tau,
\upsilon)=\underset{\{M_i\}}{\textrm{min}}\underset{i\geq
0}{\sum}\sqrt{\textrm{Tr}\{ M_i\tau\}}\sqrt{\textrm{Tr}\{
M_i\upsilon\}}\label{operational}
\end{equation}
between the statistical distributions generated by all possible
measurements of the form $M_0=P_\mathcal{K}$, $M_i=M_i^A\otimes
I^B$ for $i\geq 1$, $\underset{i\geq 0}{\sum}M_i=I^S$.

\textit{Proof.} The proof follows from Eq.~\eqref{easyform} and
Eq.~\eqref{fA}.

Note that the measure $F^A$ compares the information stored in
subsystem $\mathcal{H}^A$, which is the information extractable
through local measurements on $\mathcal{H}^A$. The last result
reflects the intuition that extracting information encoded in
$\mathcal{H}^A$ involves a measurement that projects on the
subspaces $\mathcal{H}^A\otimes\mathcal{H}^B$ or $\mathcal{K}$.

\textit{Property 2 (Normalization).} From the definition
\eqref{measure} it is obvious that
\begin{equation} F^A(\tau,
\upsilon)\leq F^A(\tau, \tau)=1 , \hspace{0.2cm} \tau \neq
\upsilon.
\end{equation}
From Eq.~\eqref{easyform} we can now see that
\begin{gather}
F^A(\tau, \upsilon)=1, \textrm{  iff
}\hspace{0.1cm}\textrm{Tr}_B\{ \mathcal{P}^{AB}(\tau) \} =
\textrm{Tr}_B\{ \mathcal{P}^{AB}(\upsilon) \},
\end{gather}
as one would expect from a measure that compares only the encoded
information in $\mathcal{H}^A$.

\textit{Proposition.} Using that the maximum in
Eq.~\eqref{measure} is attained for states of the form
$\Pi(\tau^*)$ and $\Pi(\upsilon^*)$ (Eq.~\eqref{pistar}) where
$\tau^*$ and $\upsilon^*$ satisfy Eq.~\eqref{tauupsstar} and
Eq.~\eqref{tauupsstar2}, without loss of generality we can assume
that for all $\tau$ and $\upsilon$,
\begin{equation}
F^A(\tau, \upsilon)=F(\tau^*, \upsilon^*),\label{useful}
\end{equation}
where
\begin{eqnarray}
\tau^*&=&\textrm{Tr}_B\{ \mathcal{P}^{AB}(\tau)\}\otimes
|0^B\rangle\langle 0^B|\notag\\ &+& \textrm{Tr}\{
\mathcal{P}_{\mathcal{K}}(\tau)\}|0_{\mathcal{K}}\rangle\langle
0_{\mathcal{K}}|,\label{taustar1}
\end{eqnarray}
\begin{eqnarray}
\upsilon^*&=&\textrm{Tr}_B\{ \mathcal{P}^{AB}(\upsilon)\}\otimes
|0^B\rangle\langle 0^B|\notag\\ &+& \textrm{Tr}\{
\mathcal{P}_{\mathcal{K}}(\upsilon)\}
|0_{\mathcal{K}}\rangle\langle 0_{\mathcal{K}}|,\label{upsstar1}
\end{eqnarray}
with $|0^B\rangle$ and $|0_{\mathcal{K}}\rangle$ being some fixed
states in $\mathcal{H}^B$ and $\mathcal{K}$, respectively.

\textit{Property 3 (Strong concavity and concavity of the square
of $F^A$).} The form of $F^A$ given by
Eqs.~\eqref{useful}--\eqref{upsstar1} can be used for deriving
various useful properties of $F^A$ from the properties of the
standard fidelity. For example, it implies that for all mixtures
$\underset{i}{\sum}p_i\tau_i$ and
$\underset{i}{\sum}q_i\upsilon_i$ we have
\begin{gather}
F^A(\underset{i}{\sum}p_i\tau_i,
\underset{i}{\sum}q_i\upsilon_i)=F(\underset{i}{\sum}p_i\tau^*_i,
\underset{i}{\sum}q_i\upsilon^*_i).
\end{gather}
This means that the property of \textit{strong concavity} of the
fidelity \cite{NielsenChuang00} (and all weaker concavity
properties that follow from it) as well as the \textit{concavity
of the square of the fidelity} \cite{Uhl76}, are automatically
satisfied by the measure $F^A$.

\textit{Definition 2.} Similarly to the concept of angle between
two states \cite{NielsenChuang00} which can be defined from the
standard fidelity, we can define an \textit{angle between the
encoded information in two states}:
\begin{equation}
\Lambda^A(\tau, \upsilon) \equiv \arccos F^A(\tau, \upsilon).
\end{equation}

\textit{Property 4 (Triangle inequality).} From
Eqs.~\eqref{useful}--\eqref{upsstar1} it follows that just as the
angle between states satisfies the triangle inequality, so does
the angle between the encoded information:
\begin{equation}
\Lambda^A(\tau, \upsilon) \leq \Lambda^A(\tau, \phi) +
\Lambda^A(\phi, \upsilon).
\end{equation}

\textit{Property 5 (Monotonicity of $F^A$ under local CPTP maps).}
We point out that the monotonicity under CPTP maps of the standard
fidelity does not translate directly to the measure $F^A$. Rather,
as can be seen from Eq.~\eqref{easyform}, $F^A$ satisfies
monotonicity under local CPTP maps on $\mathcal{H}^A$:
\begin{equation}
F^A(\mathcal{E}(\tau),\mathcal{E}(\upsilon) )\geq
F^A(\tau,\upsilon)
\end{equation}
for
\begin{gather}
\mathcal{E}=\mathcal{E}^A\otimes \mathcal{E}^B \oplus
\mathcal{E}_{\mathcal{K}},\label{localCPTPmaps}
\end{gather}
where $\mathcal{E}^A$, $\mathcal{E}^B$ and
$\mathcal{E}_{\mathcal{K}}$ are CPTP maps on operators over
$\mathcal{H}^A$, $\mathcal{H}^B$ and $\mathcal{K}$, respectively.

\textit{Comment.} There exist other maps under which $F^A$ is also
non-decreasing. Such are the maps which take states from
$\mathcal{H}^A\otimes\mathcal{H}^B$ to $\mathcal{K}$ without
transfer in the opposite direction. But in general, maps which
couple states in $\mathcal{H}^{A}\otimes \mathcal{H}^B$ with
states in $\mathcal{K}$, or states in $\mathcal{H}^A$ with states
in $\mathcal{H}^B$, do not obey this property. For example, a
unitary map which swaps the states in $\mathcal{H}^A$ and
$\mathcal{H}^B$ (assuming both subsystems are of the same
dimension) could both increase or decrease the measure depending
on the states in $\mathcal{H}^B$. Similarly, a unitary map
exchanging states between $\mathcal{H}^{A}\otimes \mathcal{H}^B$
and $\mathcal{K}$ could give rise to both increase or decrease of
the measure depending on the states in $\mathcal{K}$.

Finally, the monotonicity of $F^A$ under local CPTP maps implies

\textit{Property 6 (Contractivity of the angle under local CPTP
maps).} For CPTP maps of the form \eqref{localCPTPmaps},
$\Lambda^A$ satisfies
\begin{equation}
\Lambda^A(\mathcal{E}(\tau),\mathcal{E}(\upsilon)
)\leq\Lambda^A(\tau,\upsilon).
\end{equation}

\section{Robustness of OQEC with respect to initialization
errors}

Let us now consider the fidelity between the encoded information
in an ideally prepared state \eqref{perfectini} and in a state
which is not perfectly initialized \eqref{imperfect}:
\begin{gather}
F^A(\rho,\tilde{\rho})=\sqrt{\textrm{Tr}\rho_1}
\sqrt{\textrm{Tr}\tilde{\rho}_1}
F(\rho^A,\tilde{\rho}^A)+0\\
=\textrm{Tr}\sqrt{\sqrt{\textrm{Tr}_B\rho_1}\textrm{Tr}_B\tilde{\rho}_1\sqrt{\textrm{Tr}_B\rho_1}}\equiv
\check{F}(\textrm{Tr}_B\rho_1, \textrm{Tr}_B\tilde{\rho}_1).\notag
\end{gather}
After the noise process $\mathcal{E}$ with Kraus operators
\eqref{Krausoperatorsblock}, the imperfectly encoded state
transforms to $\mathcal{E}(\tilde{\rho})$. Its fidelity with the
perfectly encoded state becomes
\begin{gather}
F^A(\rho,\mathcal{E}(\tilde{\rho}))=\check{F}(\textrm{Tr}_B\rho_1,
\textrm{Tr}_B\tilde{\rho}'_1)\notag\\
=\check{F}(\textrm{Tr}_B\rho_1,
\textrm{Tr}_B\tilde{\rho}_1+\textrm{Tr}_B\{\underset{i}{\sum}D_i\tilde{\rho}_3D_i^{\dagger}\}),
\end{gather}
where we have used the expressions for $\textrm{Tr}_B\rho_1'$ and
$\textrm{Tr}_B\tilde{\rho}_1'$ obtained in Eq.~\eqref{reducedrho}
and Eq.~\eqref{reducedrhotilde}. As we pointed out earlier, the
operator
$\textrm{Tr}_B\{\underset{i}{\sum}D_i\tilde{\rho}_3D_i^{\dagger}
\}$ is positive. Then from the concavity of the \textit{square} of
the fidelity \cite{Uhl76}, it follows that
\begin{widetext}
\begin{gather}
\check{F}^2\left(\textrm{Tr}_B\rho_1,\textrm{Tr}_B\tilde{\rho}_1+\textrm{Tr}_B\{\underset{i}{\sum}D_i\tilde{\rho}_3D_i^{\dagger}
\}\right)=\label{argumentfidelity}\\
\textrm{Tr}\rho_1\textrm{Tr}\{\tilde{\rho}_1+
\underset{i}{\sum}D_i\tilde{\rho}_3D_i^{\dagger} \}
F^2\left(\rho^A,
\frac{\textrm{Tr}\tilde{\rho}_1}{\textrm{Tr}\{\tilde{\rho}_1+
\underset{i}{\sum}D_i\tilde{\rho}_3D_i^{\dagger} \}}
\tilde{\rho}^A+
\frac{\textrm{Tr}\{\underset{i}{\sum}D_i\tilde{\rho}_3D_i^{\dagger}
\}}{\textrm{Tr}\{\tilde{\rho}_1
+\underset{i}{\sum}D_i\tilde{\rho}_3D_i^{\dagger}\}}\frac{\textrm{Tr}_B\{\underset{i}{\sum}D_i\tilde{\rho}_3D_i^{\dagger}\}
}
{\textrm{Tr}\{\underset{i}{\sum}D_i\tilde{\rho}_3D_i^{\dagger}\}}
\right)\geq\notag\\
 \textrm{Tr}\rho_1\textrm{Tr}\tilde{\rho}_1
F^2(\rho^A,\tilde{\rho}^A)+\textrm{Tr}\rho_1\textrm{Tr}\{\underset{i}{\sum}D_i\tilde{\rho}_3D_i^{\dagger}
\}
 F^2\left(\rho^A,
\frac{\textrm{Tr}_B\{\underset{i}{\sum}D_i\tilde{\rho}_3D_i^{\dagger}\}
}{\textrm{Tr}\{\underset{i}{\sum}D_i\tilde{\rho}_3D_i^{\dagger}\}}\right)=\notag\\\check{F}^2(\textrm{Tr}_B\rho_1,
\textrm{Tr}_B\tilde{\rho}_1)+ \check{F}^2(\textrm{Tr}_B\rho_1,
\textrm{Tr}_B\{\underset{i}{\sum}D_i\tilde{\rho}_3D_i^{\dagger}\}
)\geq \check{F}^2(\textrm{Tr}_B\rho_1,
\textrm{Tr}_B\tilde{\rho}_1).\notag
\end{gather}
\end{widetext}
(Here, the transition from the first to the second line is
obtained by pulling out the normalization factors of the operators
in $\check{F}$ so that the latter can be expressed in terms of the
fidelity $F$. The transition form the second to the third line is
by using the concavity of the square of the fidelity. The last
line is obtained by expressing the quantities again in terms of
$\check{F}$). Therefore, we can state the following

\textit{Theorem 3.} The fidelity between the encoded information
in a perfectly initialized state \eqref{perfectini} and an
imperfectly initialized state \eqref{imperfect} does not decrease
under CPTP maps $\mathcal{E}$ with Kraus operators of the form
\eqref{Krausoperatorsblock}:
\begin{equation}
F^A(\rho,\mathcal{E}(\tilde{\rho}))\geq F^A(\rho,\tilde{\rho}).
\end{equation}

We see that even if the ``initialization-free" constraint
\eqref{extraconstraint} is not satisfied, no further decrease in
the fidelity occurs as a result of the process. The effective
noise (the term
$\textrm{Tr}_B\{\underset{i}{\sum}D_i\tilde{\rho}_3D_i^{\dagger}
\}$) that arises due to violation of that constraint, can only
decrease the initialization error.

The above result can be generalized to include the possibility for
information processing on the subsystem. Imagine that we want to
perform a computational task which ideally corresponds to applying
the CPTP map $\mathcal{C}^A$ on the encoded state. In general, the
subsystem $\mathcal{H}^A$ may consist of many subsystems encoding
separate information units (e.g. qubits), and the computational
process may involve many applications of error correction. The
noise process itself generally acts continuously during the
computation. Let us assume that all operations following the
initialization are performed fault-tolerantly
\cite{Sho96,ABO98,Kit97,KLZ98,Got97} so that the overall
transformation $\mathcal{C}$ on a \textit{perfectly initialized}
state succeeds with an arbitrarily high probability (for a model
of fault-tolerant quantum computation on subsystems, see e.g.
\cite{Ali07}). This means that the effect of $\mathcal{C}$ on the
reduced operator of a perfectly initialized state is
\begin{equation}
\textrm{tr}_B\rho_1\rightarrow
\mathcal{C}^A(\textrm{Tr}_B\rho_1)\label{FTcomputation}
\end{equation}
up to an arbitrarily small error.

\textit{Theorem 4.} Let $\mathcal{C}$ be a CPTP map whose effect
on reduced operator of every perfectly initialized state
\eqref{perfectini} is given by Eq.~\eqref{FTcomputation} with
$\mathcal{C}^A$ being a CPTP map on $\mathcal{B}(\mathcal{H}^A)$.
Then the fidelity between the encoded information in a perfectly
initialized state \eqref{perfectini} and an imperfectly
initialized state \eqref{imperfect} does not decrease under
$\mathcal{C}$:
\begin{equation}
F^A(\mathcal{C}(\rho),\mathcal{C}(\tilde{\rho}))\geq
F^A(\rho,\tilde{\rho}).
\end{equation}

\textit{Proof.} From Eq.~\eqref{FTcomputation} it follows that the
map $\mathcal{C}$ has Kraus operators with vanishing lower left
blocks, similarly to \eqref{Krausoperatorsblock}. If the state is
not perfectly initialized, an argument similar to the one
performed earlier shows that the reduced operator on the subsystem
transforms as $\textrm{Tr}_B\tilde{\rho_1}\rightarrow
\mathcal{C}^A(\textrm{Tr}_B\tilde{\rho}_1)+
\tilde{\rho}^A_{\textrm{err}}$, where
$\tilde{\rho}^A_{\textrm{err}}$ is a positive operator which
appears as a result of the possibly non-vanishing upper right
blocks of the Kraus operators. Using an argument analogous to
\eqref{argumentfidelity} and the monotonicity of the fidelity
under CPTP maps \cite{BCF96}, we obtain
\begin{gather}
F^A(\mathcal{C}(\rho),\mathcal{C}(\tilde{\rho}))=
\check{F}(\mathcal{C}^A(\textrm{Tr}_B\rho_1),\mathcal{C}^A(\textrm{Tr}_B\tilde{\rho}_1)+
\tilde{\rho}^A_{\textrm{err}})+0\notag\\
\geq
\check{F}(\mathcal{C}^A(\textrm{Tr}_B\rho_1),\mathcal{C}^A(\textrm{Tr}_B\tilde{\rho}_1))\notag\\
=\sqrt{\textrm{Tr}\rho_1}\sqrt{\textrm{Tr}\tilde{\rho}_1}
F(\mathcal{C}^A(\rho^A),\mathcal{C}^A(\tilde{\rho}^A))\notag\\
\geq\sqrt{\textrm{Tr}\rho_1}\sqrt{\textrm{Tr}\tilde{\rho}_1}
F(\rho^A,\tilde{\rho}^A)\\=\check{F}(\textrm{Tr}_B\rho_1,
\textrm{Tr}_B\tilde{\rho}_1)=F^A(\rho,\tilde{\rho}).\notag
\end{gather}

Again, the preparation error is not amplified by the process. The
problem of how to deal with preparation errors has been discussed
in the context of fault-tolerant computation on standard
error-correction codes, e.g., in \cite{Pre99}. The situation for
general OQEC is similar---if the initial state is known, the error
can be eliminated by repeating the encoding. If the state to be
encoded is unknown, the preparation error generally cannot be
corrected. Nevertheless, encoding would still be worthwhile as
long as the initialization error is smaller than the error which
would result from leaving the state unprotected.

\section{Conclusion}

In summary, we have shown that a noiseless subsystem is robust
against initialization errors without the need for modification of
the noiseless subsystem conditions. Similarly, we have argued that
general OQEC codes are robust with respect to imperfect
preparation in their standard form. This property is compatible
with fault-tolerant methods of computation, which is essential for
reliable quantum information processing. In order to rigorously
prove our result, we introduced a measure of the fidelity
$F^A(\tau, \upsilon)$ between the encoded information in two
states. The measure is defined as the maximum of the fidelity
between all possible states which have the same reduced operators
on the subsystem code as the states being compared. We derived a
simple form of the measure and discussed many of its properties.
We also gave an operational interpretation of the quantity.

Since the concept of encoded information is central to quantum
information science, the fidelity measure introduced in this paper
may find various applications. It provides a natural means for
extending key concepts such as the fidelity of a quantum channel
\cite{KL96} or the entanglement fidelity \cite{Sch96b} to the case
of subsystem codes.

\section*{Acknowledgments}

The author would like to thank Todd A. Brun for helpful
discussions and his reading of the manuscript, Daniel Lidar for
stimulating coversations, and Alireza Shabani for discussions and
pointing out the necessity for a rigorous definition of the
quantity $F^A(\tau, \upsilon)$. This research was supported in
part by NSF grant No. EMT-0524822.

\end{document}